\RequirePackage{fix-cm}
\documentclass[twocolumn]{svjour3}          % twocolumn
\smartqed  % flush right qed marks, e.g. at end of proof
\usepackage{graphicx}

\usepackage[round,sort,comma,authoryear]{natbib}
\usepackage{breakcites}
\usepackage{amsfonts}
\usepackage{amsmath}
\usepackage{graphicx}
\begin{document}

\title{Implicit Hamiltonian Monte Carlo for Sampling Multiscale Distributions%\thanks{Grants or other notes
}


\author{Arya A. Pourzanjani         \and
        Linda R. Petzold %etc.
}


\institute{Arya A. Pourzanjani \at
              Department of Computer Science\\
              University of California, Santa Barbara\\
              \email{arya@ucsb.edu}           %  \\
           \and
           Linda R. Petzold \at
           Department of Computer Science\\
              University of California, Santa Barbara\\
              \email{petzold@engineering.ucsb.edu}   
}

\date{Received: date / Accepted: date}

\maketitle

\begin{abstract}
Hamiltonian Monte Carlo (HMC) has been widely adopted in the statistics community because of its ability to sample high-dimensional distributions much more efficiently than other Metropolis-based methods. Despite this, HMC often performs sub-optimally on distributions with high correlations or marginal variances on multiple scales because the resulting stiffness forces the leapfrog integrator in HMC to take an unreasonably small stepsize. We provide intuition as well as a formal analysis showing how these multiscale distributions limit the stepsize of leapfrog and we show how the implicit midpoint method can be used, together with Newton-Krylov iteration, to circumvent this limitation and achieve major efficiency gains. Furthermore, we offer practical guidelines for when to choose between implicit midpoint and leapfrog and what stepsize to use for each method, depending on the distribution being sampled. Unlike previous modifications to HMC, our method is generally applicable to highly non-Gaussian distributions exhibiting multiple scales. We illustrate how our method can provide a dramatic speedup over leapfrog in the context of the No-U-Turn sampler (NUTS) applied to several examples.
\end{abstract}

%%%%%%%%%%%%%%%%%%%%%%%%%
%%%%%%%%%%%%%%%%%%%%%%%%%
%%%%%%%%%%%%%%%%%%%%%%%%%
\section{Introduction}
\label{intro}
% the success and advantages of HMC
The Hamiltonian Monte Carlo (HMC) algorithm \citep{duane1987hybrid} and its recent successor, the No-U-Turn (NUTS) sampler  \citep{hoffman2014no}, have seen widespread use recently in the statistics community because of their proficiency in sampling high-dimensional distributions. In fact, \cite{beskos2013optimal} showed that as the dimension, $D$, of the distribution $p(q)$ being sampled tends to infinity, HMC requires only $\mathcal{O}(D^{5/4})$ samples to sufficiently explore the distribution, while the classic random-walk Metropolis algorithm~\citep{metropolis1953equation} requires $\mathcal{O}(D^2)$. Roughly speaking, HMC and NUTS achieve this efficiency gain because rather than exploring parameter space in a random fashion, they systematically explore level-sets of Hamiltonian energy by using the leapfrog integrator to numerically simulate Hamiltonian dynamics over the potential energy surface defined by $U(q) = -\log p(q)$ \citep{neal2011mcmc}. While the Hamiltonian energy remains roughly constant over these level sets, when simulating Hamiltonian dynamics using the leapfrog integrator, the log-probability density and values of the random variables $q$ often vary quite widely in practice, yielding samples that are much closer to independent than samples from the random-walk Metropolis algorithm \citep{betancourt2017conceptual}.

% the leapfrog integrator but how it can be studied more using techniques from numerical integration
The key to HMC reaching the correct stationary distribution lies in the reversibility of the leapfrog integrator that allows the Markov transitions to satisfy detailed-balance. Furthermore, the leapfrog integrator is volume-preserving in phase-space, which makes the computation of the Metropolis probabilities in HMC trivial \citep{neal2011mcmc}. In fact, the leapfrog integrator belongs to a class of numerical integrators known as symplectic integrators that exhibit both of these properties along with a more general property known as symplecticity. Symplectic integrators have been well-studied \citep{leimkuhler2004simulating, hairer2006geometric}. However, little work has been done to characterize how the properties of a probability distribution relate to the properties of the associated Hamiltonian system in HMC, and how these properties make certain integrators more advantageous than others, depending on the problem.

%The properties of individual numerical integrators that make them more or less efficient on certain types of ODE systems that exhibit certain characteristics (e.g. multiple timescales) has been well-studied for decades and is used every day in practice by the engineering community (see \cite{ascher1998computer} for an accessible introduction). Despite this, little work has been done to characterize how the properties of a probability distribution relate to the properties of the associated Hamiltonian system in HMC and how these properties make certain integrators more advantageous than others depending on the problem.

% our contributions
To this end, we provide an analysis of how and why the geometry of Bayesian posterior distributions with low variance components can lead to multiscale Hamiltonian systems, i.e. systems with rapidly oscillating components that force leapfrog to take a much smaller stepsize than what would otherwise be possible with an implicit integrator. We describe the implicit midpoint integrator as an alternative to leapfrog for these types of problems, and show how to practically implement it in an HMC context. Furthermore, we offer practical guidelines on the stepsize to choose when using either leapfrog or implicit midpoint in an HMC sampler, as well as heuristics for when to choose one algorithm over the other. Finally, we compare, using practical examples, the efficiency of leapfrog NUTS (lfNUTS) with an implicit NUTS implementation we introduce called iNUTS. Code for these experiments is available as an R package that can be obtained by emailing the authors.

% preview of sections
In Section \ref{background} we provide a brief introduction to HMC and then describe the concept of numerical stability of an integrator and how it connects to the geometries of multivariate distributions in HMC. In Section \ref{implicit} we describe the symplectic implicit midpoint algorithm along with our practical custom implementation specific to HMC. We also derive the mathematical stability limit for the implicit midpoint algorithm, showing how it provides a clear advantage over leapfrog on multiscale systems. In Section \ref{experiments} we show how in practice, using real examples, our iNUTS implementation leads to less computational work per effective sample than leapfrog-based NUTS on common multiscale systems. We conclude with a summary of our contributions and future directions in Section \ref{discussion}.

% related work
\paragraph{Related Work:} While various works have explored modifications of the leapfrog integrator in HMC, the connection between posterior geometry and integrator choice as well as the multiscale problem has been sparsely examined. On the statistics side, both \cite{shahbaba2014split} and \cite{chao2015exponential} adapted the leapfrog integrator in HMC by splitting the potential energy into a Gaussian term and a non-Gaussian term, which are integrated separately. While these methods can alleviate the multiscale problem for distributions that can be well-approximated by a Gaussian, they fail to offer an efficiency gain over leapfrog for more complicated distributions, as we describe in Section \ref{background}. Our implicit midpoint-based method is able to achieve efficiency gains over leapfrog on a more general class of problems.

\cite{okudo2015hamiltonian} modify the leapfrog integrator in HMC by adding an auxiliary variable that allows for online adaptation of the stepsize. The RMHMC method of \cite{girolami2011riemann} and the SoftAbs extension introduced by \cite{betancourt2013general} both use the local Hessian of the potential energy function to adaptively change the mass matrix of HMC. While both of these methods can effectively adapt the stepsize in leapfrog to the local geometries of non-Gaussian distributions, they are still subject to using a small integrator stepsize in a multiscale problem, just as standard leapfrog is. In contrast, our implicit midpoint-based approach has much less severe stepsize limitations \citep{ascher1999some}.

%%%%%%%%%%%%%%%%%%%%%%%%%
%%%%%%%%%%%%%%%%%%%%%%%%%
%%%%%%%%%%%%%%%%%%%%%%%%%
\section{Hamiltonian Monte Carlo and Numerical Stability in a Multiscale Problem}
\label{background}
We provide a brief overview of the basic HMC algorithm and how it leads to a Hamiltonian system of ODEs. We then provide a short explanation and illustration of numerical stability of the leapfrog integrator on a linear Hamiltonian system and then extend this analysis to show how stability can be a crucial bottleneck in multiscale systems. Finally, using the linearization of arbitrary Hamiltonian systems, we extend the multiscale concept to non-Gaussian distributions and draw the connection between posterior geometry and the mass matrix of HMC. For a more comprehensive review of HMC see \cite{neal2011mcmc}, or \cite{betancourt2017conceptual} for a more recent review that includes an exposition on NUTS. For a more thorough review of stability analysis for the numerical solution of Hamiltonian systems see \cite[Ch. 2.6]{leimkuhler2004simulating}.

%%%%%%%%%%%%%%%%%%%%%%%%%
%%%%%%%%%%%%%%%%%%%%%%%%%
\subsection{Hamiltonian Monte Carlo}
\label{hmc}
\begin{figure}
  \includegraphics[width=0.47\textwidth]{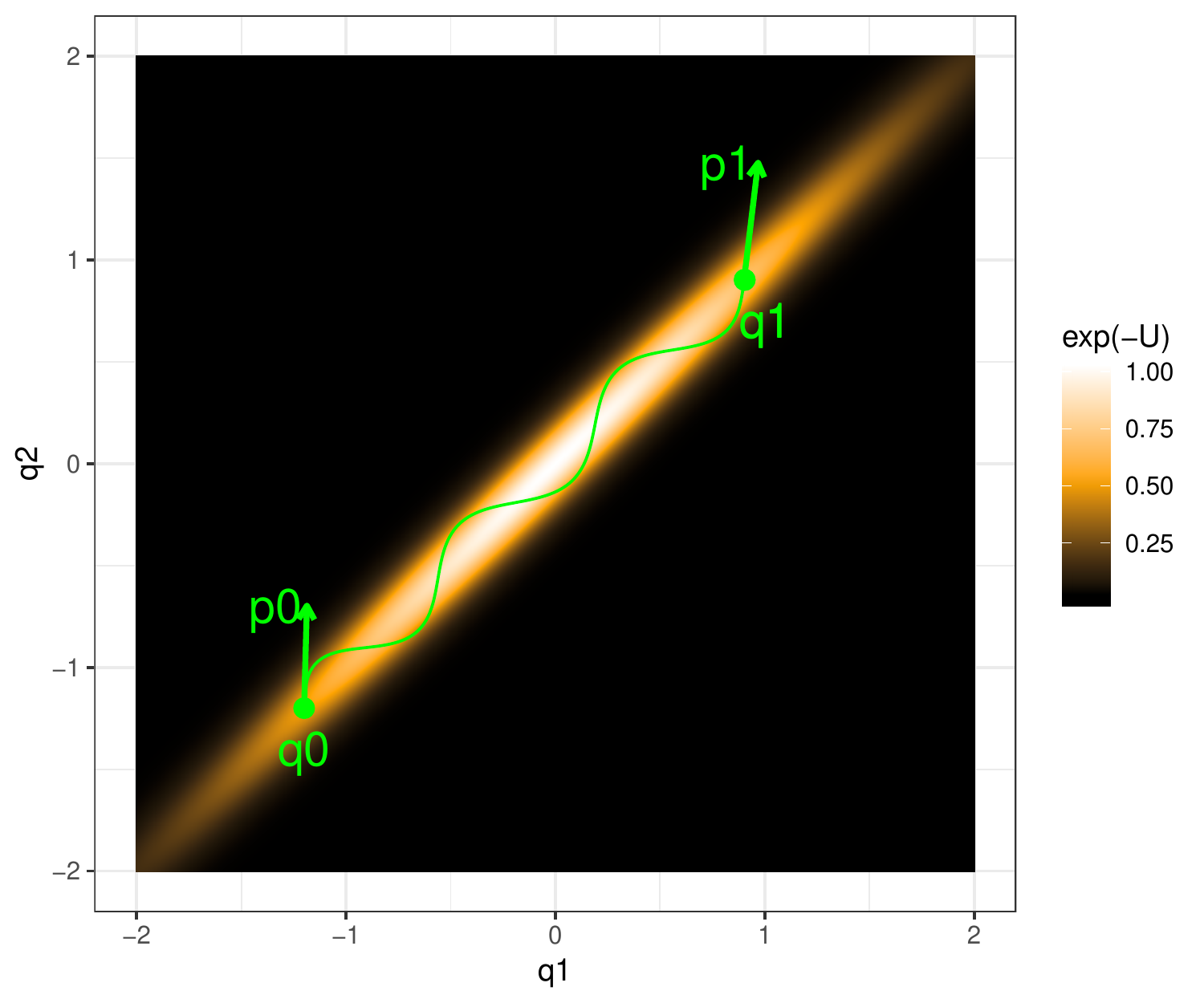}
\caption{Given a starting point $q_0$, an HMC proposal starts by drawing a momentum, $p_0$, then simulating Hamiltonian dynamics for a fixed time to reach a new proposal, $(q_1, p_1)$. A Metropolis accept-reject step is then used to accept the new proposal randomly, depending on the difference between the Hamiltonian at the original point in $q$-$p$ space and the Hamiltonian at the proposed point.}
\label{fig:hmc_proposal}       % Give a unique label
\end{figure}

HMC provides samples from an arbitrary distribution over $q \in \mathbb{R}^D$ with density $p(q)$ by taking Markov transitions that satisfy detailed-balance with respect to $p(q)$. In HMC, a potential energy surface $U(q) = -\log p(q)$ is defined, along with a kinetic energy, $K(p) :=  p^T M^{-1} p$, and a Hamiltonian, $H(q,p) := U(q) + K(p)$. Given a starting point $q_0$ on this surface, an HMC transition begins by sampling a random momentum, $p_0 \in \mathbb{R}^D$, from a multivariate normal distribution with mean zero and covariance $M$ \citep{betancourt2017conceptual}. Given these initial values $q_0$ and $p_0$, the Hamiltonian system

\begin{eqnarray}
\label{eq:hamiltonian_system}
q' &=& \frac{\partial H}{\partial p} = M^{-1}\\
p' &=& \frac{\partial H}{\partial p} =-\nabla_q\, U(q)\nonumber
\end{eqnarray}

\noindent is solved, resulting in a new set of points $q_1$ and $p_1$ in phase-space, as shown in Figure \ref{fig:hmc_proposal}. The point $q_1$ is then accepted or rejected as a new sample of the distribution, in typical Metropolis fashion using the acceptance probability defined by

\begin{equation}
\label{eq:accept_prob}
\min \left\{ 1,\; \exp \left( H(q_0, p_0) - H(q_1, p_1) \right) \right\}.
\end{equation}

In classical HMC, the dynamics defined in (\ref{eq:hamiltonian_system}) are simulated by applying discrete steps of the leapfrog integrator for $T$ steps. These leapfrog steps are discretized by a stepsize $h$, yielding

\begin{eqnarray}
\label{eq:leapfrog_update_equations}
q_{n+1} &=& q_n +  h M^{-1} p_n -  \frac{h^2}{2} M^{-1} \nabla_q\, U(q_n)   \\
p_{n+1} &=& p_n - \frac{h}{2} \nabla_q\, U(q_{n}) - \frac{h}{2} \nabla_q\, U(q_{n+1}). \nonumber
\end{eqnarray}

These update equations crucially provide a reversible and volume preserving transition due to the symplectic property of the leapfrog integrator \citep{neal2011mcmc}. Moreover, for a satisfactory stepsize the Hamiltonian, $H(q,p)$, remains nearly constant over the numerical simulation of the Hamiltonian dynamics.

%\begin{eqnarray}
%\label{eq:leapfrog_update_equations}
%p_{n+1/2} &=& p_n - \frac{h}{2} \nabla_q\, U(q_n) \nonumber\\
%q_{n+1} &=& q_n + h M^{-1} p_{n+1/2} \\
%p_{n+1} &=& p_n - \frac{h}{2} \nabla_q\, U(q_{n+1}) \nonumber
%\end{eqnarray}

%%%%%%%%%%%%%%%%%%%%%%%%%
%%%%%%%%%%%%%%%%%%%%%%%%%
\subsection{Numerical Stability}
\label{numerical_stability}
In practice, the stepsize of the leapfrog integrator is limited by its numerical stability, which is problem-dependent. This concept is easiest to illustrate using a simple system derived from a univariate Gaussian. Specifically, for a univariate Gaussian distribution with mean zero and variance $\sigma^2$, the potential energy function used in HMC is $U(q) = q^T \Sigma^{-1} q$. For an identity mass matrix, this leads to the following leapfrog update rule:

\begin{eqnarray}
\label{eq:leapfrog_update_equations_gaussian}
q_{n+1} &=& q_n +  h p_n -  \frac{h^2}{2} \sigma^2 q_n   \\
p_{n+1} &=& p_n - \frac{h}{2} \sigma^2 q_{n} - \frac{h}{2} \sigma^2 q_{n+1}, \nonumber
\end{eqnarray}

\noindent which can be compactly written in matrix form as

\begin{equation}
\label{eq:leapfrog_update_matrix_gaussian}
\begin{pmatrix}
q_{n+1}\\
p_{n+1}
\end{pmatrix}
=
\begin{pmatrix}
1 - \frac{h^2}{2} \sigma^2 & h\\
-h \sigma^2 (1 - \frac{h^2}{4} \sigma^2) & 1 - \frac{h^2}{2} \sigma^2
\end{pmatrix}
\begin{pmatrix}
q_{n}\\
p_{n}
\end{pmatrix}.
\end{equation}

\begin{figure*}
  \includegraphics[width=0.98\textwidth]{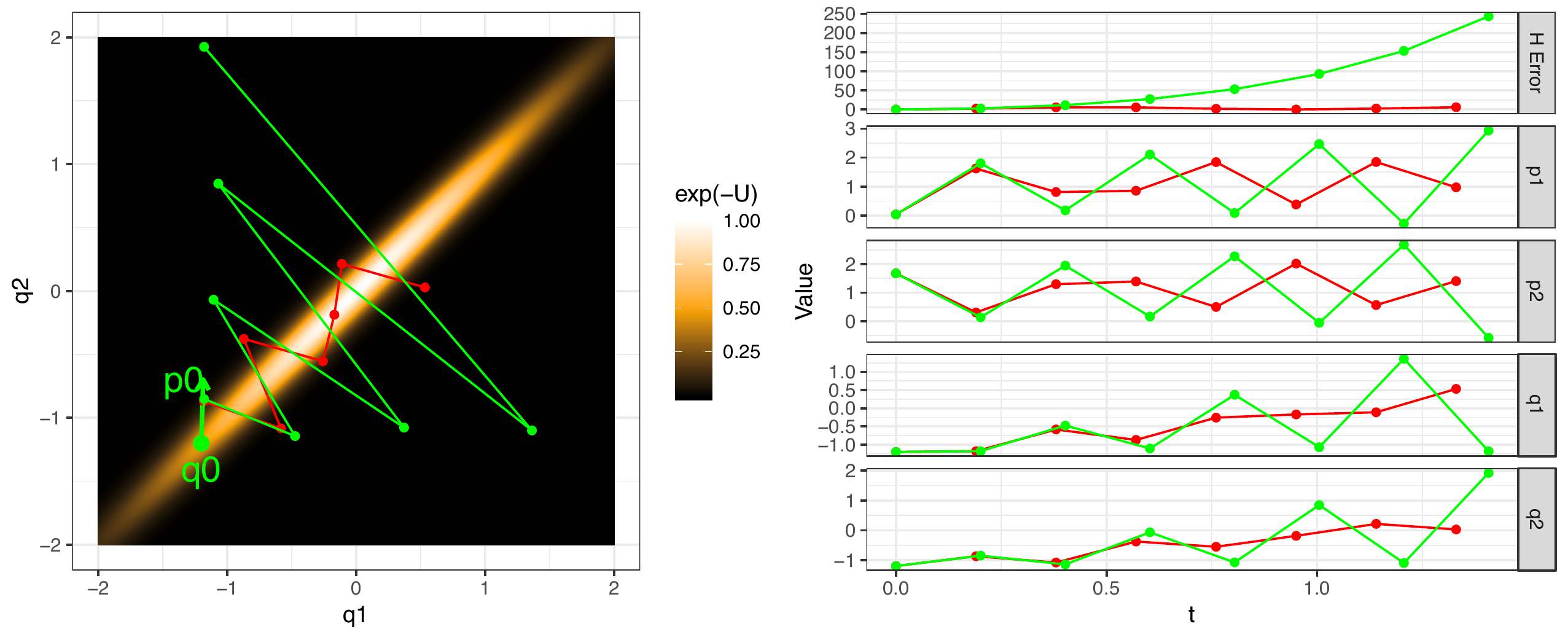}
\caption{For stepsizes larger than the stability threshold of the problem, the leapfrog method goes unstable, leading to wild numerical solutions whose errors characteristically grow after each step (green trajectory). This is in contrast to numerical solutions below the stability threshold, for which the error stays bounded after a series of steps (red trajectory).}
\label{fig:instability}       % Give a unique label
\end{figure*}

The update matrix in (\ref{eq:leapfrog_update_matrix_gaussian}) describes how leapfrog advances forward one step, for a univariate Gaussian system with mass one. Because the leapfrog method is symplectic, the determinant of this matrix is always one. However, for $h < 2/\sigma$ the eigenvalues are complex and have modulus equal to one, while for $h > 2/\sigma$ they are real with one of the eigenvalues having modulus greater than one \cite[Ch. 2.6]{leimkuhler2004simulating}. The latter case results in numerical instability of the integrator \cite{leimkuhler2004simulating}. Intuitively, this means that any small numerical errors that will inevitably arise in the numerical solutions $(q_n, p_n)$ will be successively ``magnified" by each application of the leapfrog update matrix, quickly resulting in a solution with unacceptably large error (Figure \ref{fig:instability}).

The univariate analysis can be readily extended to a multivariate Gaussian of dimension $D$ with covariance matrix $\Sigma$. With a mass matrix equal to the identity, this results in the following Hamiltonian system:

\begin{eqnarray}
\label{eq:hamiltonian_system_multivariate_guassian}
q' &=& p\\
p' &=& \Sigma^{-1} q.\nonumber
\end{eqnarray}

\begin{figure*}
  \includegraphics[width=0.98\textwidth]{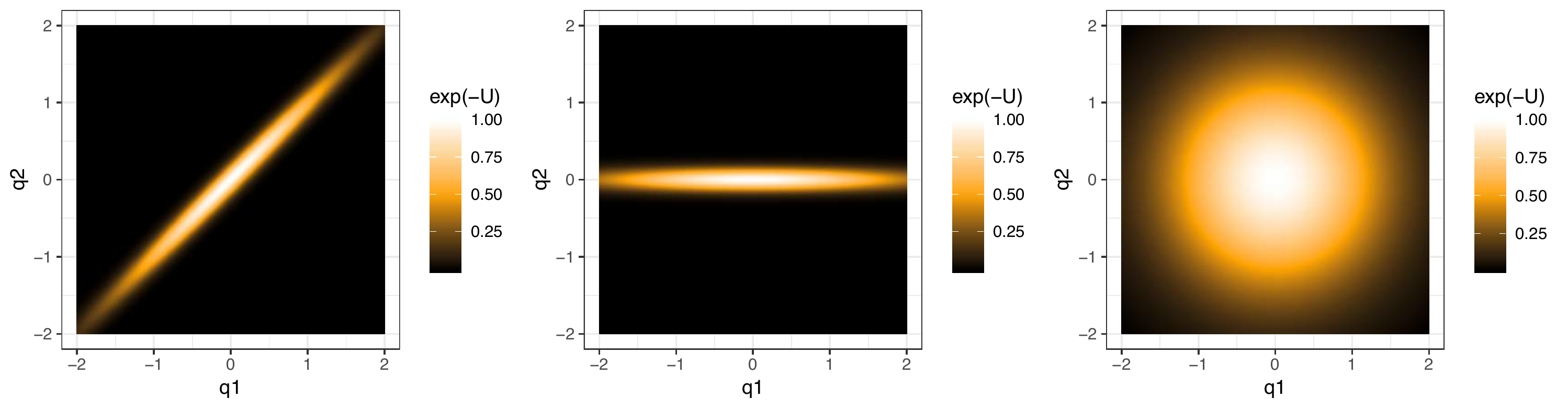}
\caption{A decoupling transformation is equivalent to transforming a correlated distribution (left) to an uncorrelated one (middle). Similarly, a transformation can be used to make an arbitrary Gaussian isotropic (right).}
\label{fig:transformation}       % Give a unique label
\end{figure*}

In general, $\Sigma^{-1}$ may have off-diagonal correlation terms that make this derived system coupled. However, the transformation defined by $u:= V^{-1} q$ and $w := V^{-1} p$, where $V$ is a matrix whose columns consist of the eigenvectors of $\Sigma^{-1}$, essentially uncouples the differential equations, yielding

\begin{eqnarray}
\label{eq:hamiltonian_system_uncoupled_multivariate_guassian}
u' &=& w\\
w' &=& \Lambda u,\nonumber
\end{eqnarray}

\noindent where $\Lambda$ is a diagonal matrix composed of the eigenvalues, $\lambda_1, \cdots, \lambda_D$, of $\Sigma^{-1}$ \cite[Ch. 2.6]{leimkuhler2004simulating}. This decoupling transformation can be thought of as reparameterizing the problem so that the potential energy surface exhibits no correlation (Figure \ref{fig:transformation}).

Applying the leapfrog update rule (\ref{eq:leapfrog_update_equations}) to the uncoupled equations (\ref{eq:hamiltonian_system_uncoupled_multivariate_guassian}) elucidates how the issue of numerical stability arises in the case of a multivariate Gaussian. Specifically, if one or more of the eigenvalues of $\Sigma^{-1}$ do not satisfy the inequality $h < 2/\sqrt{\lambda_i}$, then the leapfrog update matrix will have a real eigenvalue with modulus greater than one, and instability in the numerical integration will arise. In practice this will lead to poor acceptance rates in the Markov chain as well as to divergences \citep{betancourt2017conceptual}. Formally, for leapfrog on a multivariate Gaussian with mass matrix $I$ the stepsize $h$ must satisfy the inequality 

\begin{equation}
\label{eq:stability_definition}
h < 2 / \sqrt{\rho(\Sigma^{-1})}
\end{equation}

\noindent to ensure stability (here $\rho(\Sigma^{-1})$ denotes the spectral radius of $\Sigma^{-1}$, defined as the maximum of the absolute value of its eigenvalues).

This condition can severely limit the timestep of leapfrog. Intuitively, the dimensions of the distribution that exhibit high variance, and thus low curvature in the potential energy surface, will have slowly oscillating Hamiltonian trajectories that are very smooth and can be integrated by leapfrog with a reasonable stepsize. Meanwhile, dimensions with low variance, and thus high curvature, will lead to rapdily oscillating solutions that require a very small stepsize. Thus a system that has even one state with a dramatically smaller variance than the others will force leapfrog to take excessively small steps on the whole system. An ODE system that exhibits this sort of behavior is known as a multiscale system \citep{ascher1998computer}. In practice, the telltale sign of a distribution that will lead to a multiscale Hamiltonian system in HMC is the inverse covariance matrix, $\Sigma^{-1}$ having a large condition number. The condition number $\kappa(\Sigma^{-1})$ is defined as the ratio of the largest eigenvalue of $\Sigma^{-1}$ to the smallest. Intuitively, it captures the ratio of the curvatures of the dimensions of the potential energy surface. Practically speaking, a large condition number can arise either when an uncorrelated Gaussian has a large disparity in the variance of its largest and smallest dimensions, or alternatively when a multivariate Gaussian contains high correlations between dimensions. In engineering, the multiscale problem is often resolved by using an appropriate implicit integrator, which is typically able to take  much larger steps on multiscale problems while preserving the accuracy of desired quantities of the system \cite{ascher1999some, leimkuhler2016molecular}. We describe this approach and its application to HMC in the following section.

%%%%%%%%%%%%%%%%%%%%%%%%%
%%%%%%%%%%%%%%%%%%%%%%%%%
\subsection{Nonlinear Posterior Geometry}
\begin{figure}
  \includegraphics[width=0.47\textwidth]{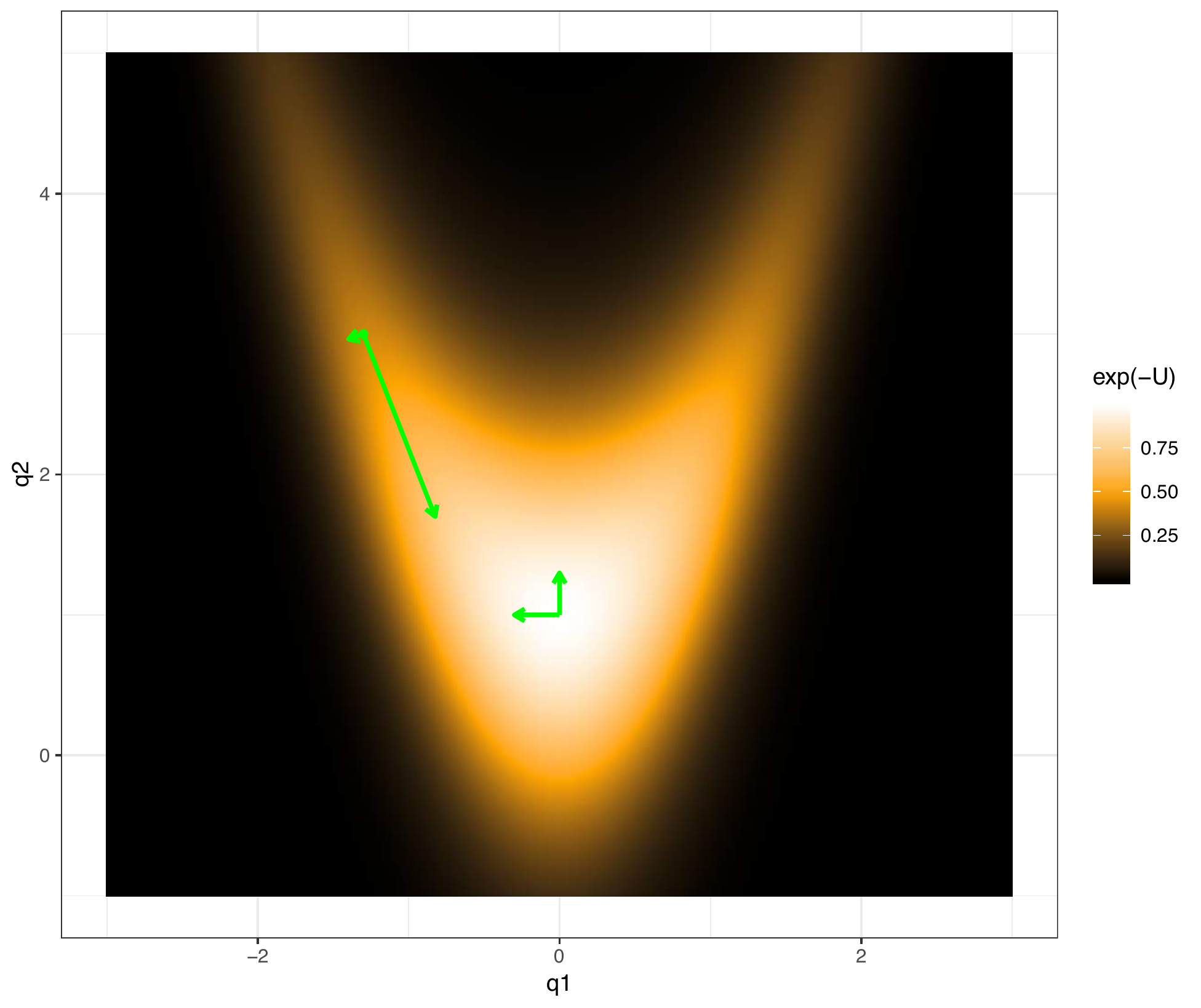}
\caption{Unlike a Gaussian distribution, arbitrary non-Gaussian distributions have non-constant local Hessians (here characterized by the green eigenvectors of the local Hessian multiplied by their associated eigenvalues) that affect the oscillatory frequency of HMC trajectories and thus an acceptable leapfrog stepsize.}
\label{fig:nonlinear}       % Give a unique label
\end{figure}

While the stability analysis considered so far has been only for Gaussian distributions which result in linear Hamiltonian systems, most practical distributions being sampled by HMC are not Gaussian and thus result in nonlinear Hamiltonian systems. For nonlinear systems, one can locally identify a multiscale system by analyzing the condition number of the local Hessian of the potential energy surface, $\nabla_{qq} U(q)$. This local Hessian is equivalent to the inverse covariance matrix of the local Laplace approximation to the posterior \citep[Chapter 4]{gelman2013bayesian}. Thus the key difference between the posterior geometry of a Gaussian distribution and that of a more complicated distribution is that the former has a potential energy surface with a constant Hessian while the latter may contain vastly different local Hessians throughout the surface, which can lead to different multiscale properties of the associated Hamiltonian system (Figure \ref{fig:nonlinear}). In practice, this means that different leapfrog stepsizes may be required, depending on where on the potential energy surface the sampler is currently located. Roughly speaking, RMHMC uses local Hessian evaluations of the potential energy surface to adaptively change this stepsize based on the local curvature \citep{betancourt2013general}. %For general, non-Hamiltonian systems, most production-quality ODE solvers use the the largest eigenvalue of the Jacobian of the system to locally select an integrator stepsize. This is equivalent to monitoring the eigenvalues of the local Hessian in the Hamiltonian case \citep{petzold1983automatic}.

%

%%%%%%%%%%%%%%%%%%%%%%%%%
%%%%%%%%%%%%%%%%%%%%%%%%%
\subsection{The Mass Matrix as a Reparameterization}
Although a multivariate Gaussian system exhibiting multiple scales can significantly hinder the efficiency of the leapfrog method, this deficiency may be resolved by appropriately selecting the mass matrix $M$ in (\ref{eq:hamiltonian_system}) \citep{roberts2002langevin, girolami2011riemann}. In fact, utilizing a mass matrix in HMC that is equal to the covariance $\Sigma$ of the Gaussian is equivalent to reparameterizing the problem to an equivalent isotropic Gaussian \citep{neal2011mcmc} (Figure \ref{fig:transformation}). From a numerical analysis perspective, $M^{-1}$ can be viewed as a preconditioner that transforms the problem so as to give $\Sigma^{-1}$ a better condition number $\kappa(\Sigma^{-1})$, i.e. it transforms the problem so that there is less discrepency between the largest and smallest eigenvalues of $\Sigma^{-1}$.

While a conveniently selected mass matrix can effectively eliminate the multiscale problem of leapfrog for Gaussian distributions, for distributions with more complicated geometry whose local curvature varies, a constant mass matrix is inherently much less effective in accounting for multiscale geometry \citep{christensen2005scaling}. Furthermore, as \cite{girolami2011riemann} point out, it is unclear how such a global mass matrix could be defined in a systematic manner. Splitting methods such as that of \cite{shahbaba2014split} as well as \cite{chao2015exponential} which can handle high curvature and multiple scales by separating out a constant Gaussian approximation of the distribution unfortunately have the same limitation, as they cannot handle the varying curvature of non-Gaussian posteriors.

%%%%%%%%%%%%%%%%%%%%%%%%%
%%%%%%%%%%%%%%%%%%%%%%%%%
%%%%%%%%%%%%%%%%%%%%%%%%%
\section{Implicit HMC}
\label{implicit}
The stability bottleneck placed on leapfrog by a multiscale system is characteristic of explicit integrators like leapfrog. In the ODE community, implicit integrators have been used to essentially ``skip" fast oscillations while accurately evolving in time quantities of interest in the system \citep{ascher1999some, leimkuhler2016molecular}. We describe the implicit midpoint integrator: a symplectic alternative to leapfrog that is of the same order of accuracy as leapfrog, but is implicit, allowing it to take much bigger timesteps on multiscale problems than what would otherwise be possible with leapfrog. Unlike the approaches of \cite{shahbaba2014split} and \cite{chao2015exponential}, the implicit midpoint integrator is applicable to arbitrary multiscale systems, not just Gaussian ones.

We explain how, unlike leapfrog, the implicit midpoint method has no stability limit on a linear system. We then describe a custom Newton-Krylov method for the solution of the nonlinear system that must be solved at each timestep by midpoint. Finally, we point out that while the implicit midpoint method is unconditionally stable on a linear system, in practice it can go unstable on nonlinear problems \citep{ascher1999some}, although at a much larger stepsize than leapfrog is able to take. We discuss the practical implications of this and aspects of choosing between leapfrog and implicit midpoint for a specific problem, and we give practical guidelines for selecting a stepsize for both methods.

%%%%%%%%%%%%%%%%%%%%%%%%%
%%%%%%%%%%%%%%%%%%%%%%%%%
\subsection{Implicit Midpoint and Stability}
For general Hamiltonian systems of the form in equation \ref{eq:hamiltonian_system}, the timesteps of implicit midpoint are defined by

\begin{eqnarray}
\label{eq:midpoint_update_equations}
q_{n+1} &=& q_n +  h M^{-1} \left( \frac{p_n + p_{n+1}}{2} \right)   \\
p_{n+1} &=& p_n - h \nabla_q\, U\left( \frac{q_n + q_{n+1}}{2} \right). \nonumber
\end{eqnarray}

Because the point $(q_{n+1},\, p_{n+1} )$ is defined implicitly, as opposed to in leapfrog where an explicitly computable update equation is given, one must resort to either functional iteration or Newton's method to compute its value. However, in situations where implicit midpoint would be desirable over leapfrog, Newton's method or a more robust modification of it is typically preferred \citep[ch. 3.4.2]{ascher1998computer}. We elaborate on this point in the subsequent subsection.

\begin{figure*}
  \includegraphics[width=0.98\textwidth]{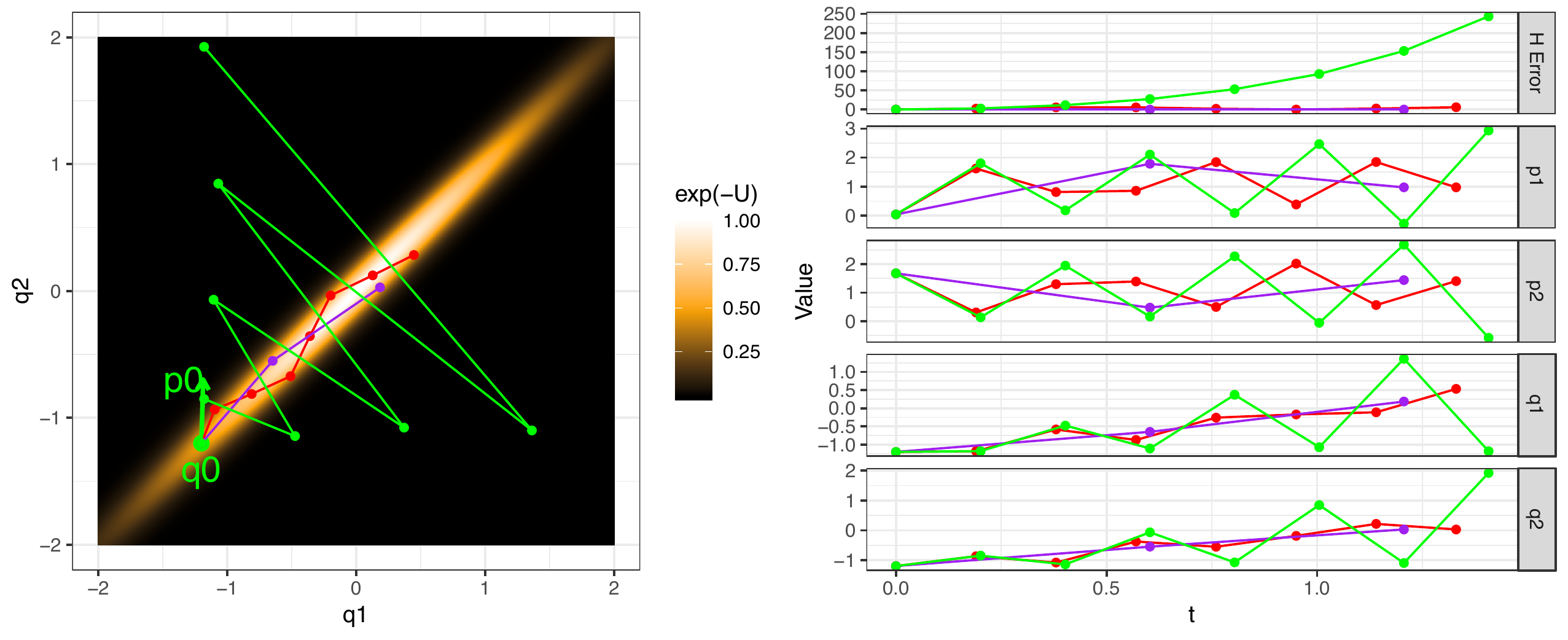}
\caption{Unlike the leapfrog integrator which goes unstable for large enough stepsizes (green trajectory), the implicit midpoint integrator can remain stable at the same stepsize (red trajectory) and even at stepsizes much larger than the stability threshold for leapfrog (purple trajectory).}
\label{fig:midpoint_stability}       % Give a unique label
\end{figure*}

An analaysis that is similar to the one performed for leapfrog in section \ref{numerical_stability} shows that unlike leapfrog, the implicit midpoint method does not have a stability limit on the linear system that arise from HMC sampling of a Gaussian distribution. In fact, like the update matrix for leapfrog, the update matrix for midpoint on a Gaussian distribution always has determinant one, however it can be shown that its eigenvalues are complex for any stepsize, $h$ (see appendix \ref{appendix_implicit_eigenvalues} for a full analysis of the stability of implicit midpoint for linear Hamiltonian systems). In other words, the implicit midpoint method is stable on Gaussian systems for any stepsize, i.e. it has not stability limit (Figure \ref{fig:midpoint_stability}).

%

%%%%%%%%%%%%%%%%%%%%%%%%%
%%%%%%%%%%%%%%%%%%%%%%%%%
\subsection{Fast Solution of the Nonlinear System}
As previously mentioned, (\ref{eq:midpoint_update_equations}) must typically be solved for $(q_{n+1}\, p_{n+1})$ numerically. In practice this can be done by applying Newton's method to solve equation \ref{eq:midpoint_update_equations}, however this can be costly for large systems.
First, because the first $D$ equations in \ref{eq:midpoint_update_equations} are linear, one can substitute the first $D$ equations into the second $D$ to obtain the nonlinear system \ref{eq:midpoint_simplified_system} with $D$ equations and the $D$ unknowns, $p_{n+1}$. Once $p_{n+1}$ is obtained, $q_{n+1}$ can be obtained by simply plugging the result back into the first set of equations, yielding

\begin{equation}
\label{eq:midpoint_simplified_system}
p_{n+1} = p_n - h \nabla_q U \left( q_n + \frac{h}{4} M^{-1} (p_n  + p_{n+1} )\right).
\end{equation}

Renaming $p_{n+1}$ to $x$ and noting that $p_n$ is known from the previous timestep, equation (10) can be written as

\begin{equation}
\label{eq:nonlinear_system_standard_form}
g(x) := x - p_n + h \nabla_q U \left( q_n + \frac{h}{4} M^{-1} (p_n  + x )\right).
\end{equation}

Newton's method applied to this system consists of starting with an initial guess, $x_0$, and iteratively obtaining approximate solutions $x_1, x_2, \cdots$ using the following steps until a convergence criteria is reached:

\begin{enumerate}
\item Solve the linear system $J_g(x_n) \delta = -g(x_n)$
\item Set $x_{n+1} = x_n + \alpha \delta$,
\end{enumerate}

\noindent where for the standard Newton iteration $\alpha =1$. Here, $J_g(x_n)$ refers to the Jacobian of $g$ evaluated at $x_n$ and is given by

\begin{equation}
\label{eq:midpoint_explicit_jacobian}
J_g(x) := I + \frac{h^2}{4} \nabla_{qq} U \left( q_n + \frac{h}{4} M^{-1} (p_n  + x )\right) M^{-1}.
\end{equation}

\noindent For large systems, exploitation of the structure of the system is critical for both efficiency and robustness. We introduce the Newton-Krylov method, which is particularly well-suited to this class of problems. %In addition, $\alpha$ is set to one. In practice, the steps of the classic Newton method can be modified to achieve better efficiency or robustness depending on the problem. We describe the modifications that we found most useful in practice and provide a brief justification of each one.

%%%%%%%%%%%%%%%%%%%%%%%%%
\subsubsection{Newton-Krylov Methods}
%A straightforward implementation of Newton's method requires computation of the Jacobian matrix and solution of a linear system at each timestep. This would be impractical for large systems. Instead, we propose using a Newton-Krylov method which can take advantage of the structure of many Bayesian problems.

In the classic Newton method, the linear solve step requires explicitly computing $J_g(x_n)$ and carrying out a full linear solve of the equation $J_g(x_n) \delta = -g(x_n)$ for $\delta$ at each Newton iteration. This requires evaluating the Hessian $\nabla_{qq} U(q)$ of the potential energy, which scales as $\mathcal{O}(D^2)$. A Newton-Krylov method \citep{knoll2004jacobian} circumvents this expensive linear solve that occurs at every Newton iteration by replacing the linear solve with an approximate linear solve that is much cheaper to compute. Specifically, the Newton-Krylov method replaces the solution $\delta$ with an approximate solution $\tilde{\delta}$ such that $\| J_g(x_n) \tilde{\delta} + g(x_n) \|$ is less than some tolerance, $\eta_n$, otherwise known as a ``forcing term". This forcing term is typically selected using a scheduling criteria that satisfies certain theoretical properties. For our experiments we found the most success with method 2.1 of \cite{eisenstat1996choosing}, as it did not rely on manually selected ``tuning parameters".  %Although it finds an approximation of $\delta$, the Newton-Krylov method still maintains fast convergence properties in practice.

In our iNUTS implementation, we use the GMRES solver \citep{saad1986gmres} to compute the approximate linear solves within the Newton-Krylov iterations. The GMRES algorithm works by iteratively computing approximate solutions to the linear system $J_g(x_n) \delta = -g(x_n)$ that reduce the norm $\| J_g(x_n) \delta + g(x_n) \|$, while needing only to evaluate the product of the matrix $J_g(x_n)$ with an arbitrary vector $v$.

Needing only to evaluate the product of the matrix $J_g(x_n)$ with an arbitrary vector $v$, rather than computing $J_g(x_n)$ and then computing its product with $v$, is particularly advantageous in HMC for two reasons. First, the Jacobian vector product $J_g(x_n) \cdot v$ can be evaluated using only a Hessian-vector product of the potential energy which, in an efficient automatic differentiation implementation such as the one available in Stan \citep{carpenter2017stan}, scales as $\mathcal{O}(D)$ as opposed to computing a full Hessian, which scales as $\mathcal{O}(D^2)$. Second, the iterated solutions in Krylov-based solver are known to converge faster when the matrix $J_g(x_n)$ has ``clusters" of eigenvalues that are close together \citep{meyer2000matrix}, as is the case when the system being solved comes from the posterior of a Bayesian model, particularly a multilevel Bayesian model. Specifically, in the case of implicit midpoint applied to HMC on the posterior of a Bayesian multilevel model, the matrix (\ref{eq:midpoint_explicit_jacobian}) will have clusters of eigenvalues that are all of similar order and correspond to the units in a multilevel model that are are all at the same level. For example, for the famous ``eight schools" model in \cite{gelman2013bayesian}, there is a variance parameter $\tau$ that is at the scale of 1, and eight separate school-level parameters at the scale of 10. Although there are nine parameters total and thus a nine-by-nine system to solve, in practice the Newton-Krylov method can typically solve the linear system to numerical precision in only two iterations, because there are only two ``clusters" of eigenvalues in the Hessian: the cluster at the scale of 1 and the cluster at the scale of 10.

%%%%%%%%%%%%%%%%%%%%%%%%%
\subsubsection{Using a Line Search}
The second modification we make to the classic Newton method is to change the steplength $\alpha$ of the Newton iteration to ensure that consecutive iterations of the Newton-Krylov method effectively reduce the residual error in the nonlinear system. In particular, we use a geometric line-search to continually halve the size of $\alpha$ until the new solution satisfies the Armijo-Goldstein condition \citep{quarteroni2010numerical}.

%%%%%%%%%%%%%%%%%%%%%%%%%
\subsubsection{Choosing an Initial Guess}
While not a modification of Newton's method per se, in our iNUTS implementation we also use a unique method of setting the initial guess $x_0$ to the nonlinear solver. In practice, this greatly improves the number of Newton-Krylov iterations needed for convergence. In particular, we exploit the numerical properties of the implicit midpoint integrator on multiscale systems, to obtain a good initial guess. Specifically, when the stepsize $h$ of the numerical integrator is small relative to the local frequency of a particular oscillating momentum coordinate $p^{(i)}$, the previous momentum, $p_{n}^{(i)}$, will serve as a good initial guess to $p_{n+1}^{(i)}$ in the system \ref{eq:midpoint_simplified_system},  as the numerical solution will not be changing much between consecutive steps. Similarly, the second to last momentum $p_{n-1}^{(i)}$ will also serve as a good initial guess, although not quite as good as the last value. On the other hand, when $h$ is much larger than the frequency of a particular oscillating variable $p^{(i)}$, which will be the case in a multiscale system, the numerical solution will approximately ``alternate" about zero taking on the values $p_1^{(i)} = \gamma, p_2^{(i)} = -\gamma, p_3^{(i)} = \gamma, \cdots$. Thus in this case, the second to last momentum serves as a good initial guess. Thus we use $p_{n-1}$ as an initial guess overall for the nonlinear equation solver.

%%%%%%%%%%%%%%%%%%%%%%%%%
%%%%%%%%%%%%%%%%%%%%%%%%%
\subsection{Nonlinear Stability Limit and Choosing an Integrator and a Stepsize}
While the implicit midpoint method is unconditionally stable for any stepsize $h$ on a linear system, it can and will exhibit instability for certain nonlinear systems, although typically at a stepsize that is much larger than the stepsize at which leapfrog would go unstable on the same problem \citep{ascher1999some}. In practice, these instabilities can typically be identified by observing large growth in the Hamiltonian of the numerical numerical trajectory, or when Newton-Krylov iterations fail to converge. To find an appropriate stepsize for implicit midpoint in practice, we recommend running a sampler ``warmup" period where the stepsize $h$ is reduced by some factor like $1/2$ until these pathological behaviors are eliminated. This is akin to the current warmup method of Stan, where the stepsize of the integrator is reduced until an acceptable accept rate is reached \citep{carpenter2017stan}.

To select between implicit midpoint and leapfrog on a specific problem, we suggest starting with implicit midpoint and periodically calculating an approximation to the largest eigenvalue of the local Hessian of the potential energy function during a warmup period. This eigenvalue approximation can be efficiently computed with only a few Hessian-vector products using the power method of numerical linear algebra \citep{meyer2000matrix}. This can in turn be used to get an approximation of the largest stepsize that leapfrog could take while maintaining stability, via equation \ref{eq:stability_definition}. When the smallest of these approximate stepsizes divided by two is close to the implicit midpoint stepsize needed to maintain stability divided by the average number of gradient evaluations plus Hessian-vector products required by implicit midpoint, then the leapfrog method will be more efficient for the problem. This is because the leapfrog method requires two gradient evaluations per step, and a Hessian-vector product, like a gradient evaluation, scales linearly in the number of inputs in an automatic differentiation system such as the one used in Stan. In practice, we found that computation of the Hessian-vector product was 1.2-1.3 times slower than a gradient evaluation in Stan.

%%%%%%%%%%%%%%%%%%%%%%%%%
%%%%%%%%%%%%%%%%%%%%%%%%%
%%%%%%%%%%%%%%%%%%%%%%%%%
\section{Experiments}
\label{experiments}
We illustrate on several examples how implicit midpoint can achieve superior efficiency over leapfrog on multiscale problems. For all of our examples we compare using a custom implementation of NUTS that either uses leapfrog (lfNUTS) or implicit midpoint (iNUTS). Our code is freely available as an R package, and at the time of writing can be obtained by emailing the authors.

%%%%%%%%%%%%%%%%%%%%%%%%%
%%%%%%%%%%%%%%%%%%%%%%%%%
\subsection{2-D Gaussian}
As discussed in Section \ref{numerical_stability}, the leapfrog stability limit for a multivariate Gaussian is directly related to the largest eigenvalue of the inverse covariance matrix. One of the ways this eigenvalue can be large is when the multivariate Gaussian is highly correlated. To illustrate how this affects NUTS in practice, we compared the average tree depth per sample of lfNUTS and iNUTS for varying levels of $\rho$ on the two-dimensional Gaussian distribution with mean zero, and covariance matrix

\begin{equation}
\Sigma
=
\begin{pmatrix}
1 & \rho \\
\rho & 1
\end{pmatrix}.
\end{equation}

\begin{figure}
  \includegraphics[width=0.47\textwidth]{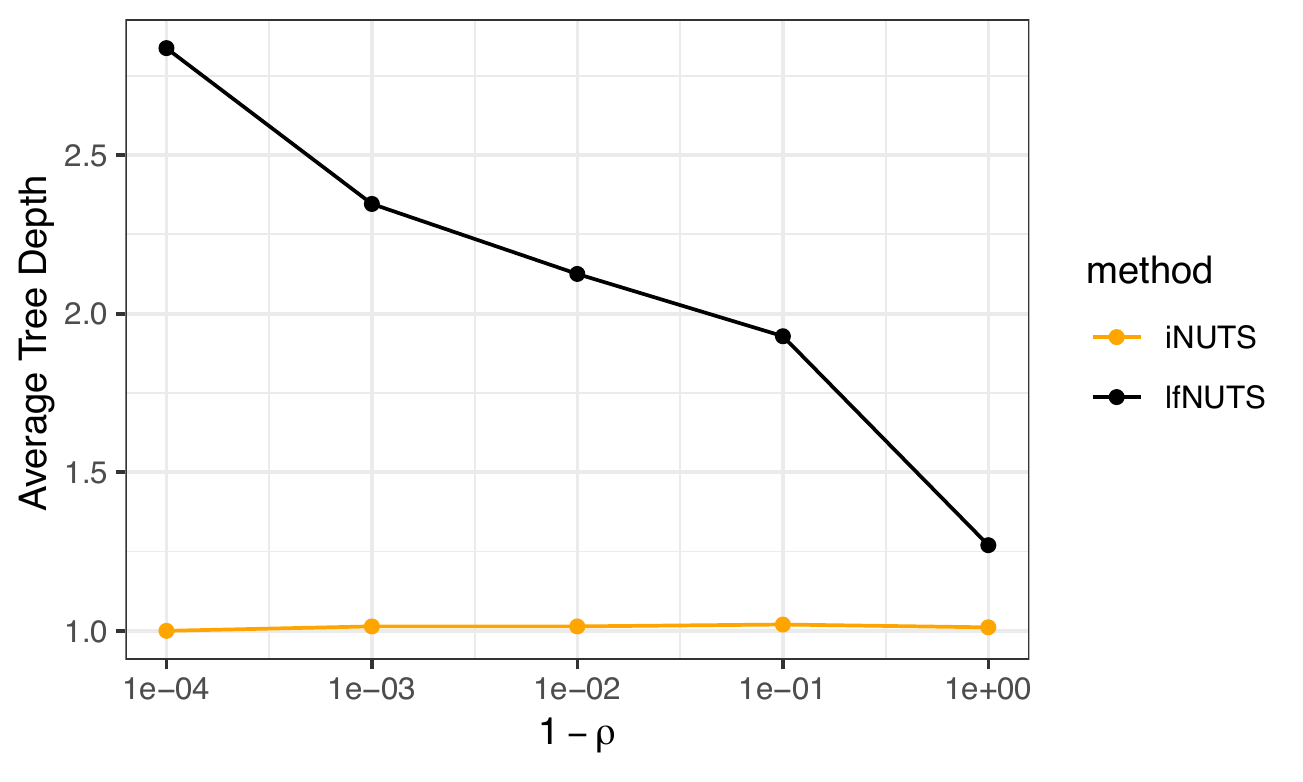}
\caption{For highly correlated Gaussian distributions, the leapfrog integrator in lfNUTS is forced to take tiny stepsizes that result in larger average NUTS tree depths. As the correlation goes to zero, the average tree depth approaches that of iNUTS.}
\label{fig:treedepth}       % Give a unique label
\end{figure}

The empirical scaling of average tree depth over 1,000 NUTS samples versus $1-\rho$ is shown in Figure \ref{fig:treedepth}. In the NUTS algorithm, tree depth is selected automatically, when a Hamiltonian trajectory has ``U-turned". Two to the power of the tree depth represents how many steps in the trajectory were computed. For more correlated distributions, leapfrog is forced to take smaller stepsizes, which results in more steps having to be taken and thus larger tree depths. Note that for a Gaussian system, implicit midpoint has no stepsize limit for stability regardless of the covariance of the Gaussian. Thus the stepsize can be chosen large enough so that a tree depth of only one is required.

%

%%%%%%%%%%%%%%%%%%%%%%%%%
%%%%%%%%%%%%%%%%%%%%%%%%%
%\subsection{250-D Gaussian}
%To compare lfNUTS and iNUTS on a larger Gaussian and illustrate how the multiscale problem negatively impacts the leapfrog integrator,  we drew a covariance matrix, $A$, from a Wishart distribution with identity scale matrix and 250 degrees of freedom and took NUTS samples from a multivariate Gaussian with mean zero and covariance $A$ as in \cite{hoffman2014no}. A summary of results is presented in Table 1.

%%%%%%%%%%%%%%%%%%%%%%%%%
%%%%%%%%%%%%%%%%%%%%%%%%%
\subsection{Banana Distribution}
To illustrate how the implicit midpoint integrator is advantageous for distributions with nonlinear correlations that cannot be addressed by a mass matrix, we compared lfNUTS and iNUTS on a highly-correlated banana distribution \citep{long2013banana} with parameters $a = 1$ and $b = 100$ whose probability density function (PDF) takes the following form: 

\begin{eqnarray}
&&p(q_1, q_2) \nonumber\\
&&=\frac{1}{2\pi} \exp \left\{- \frac{1}{2} \left[ a^2 q_1^2 + \frac{q_2 - ab(a^2 q_1^2 + a^2)}{a^2}\right] \right\}.
\end{eqnarray}

\begin{figure}
  \includegraphics[width=0.47\textwidth]{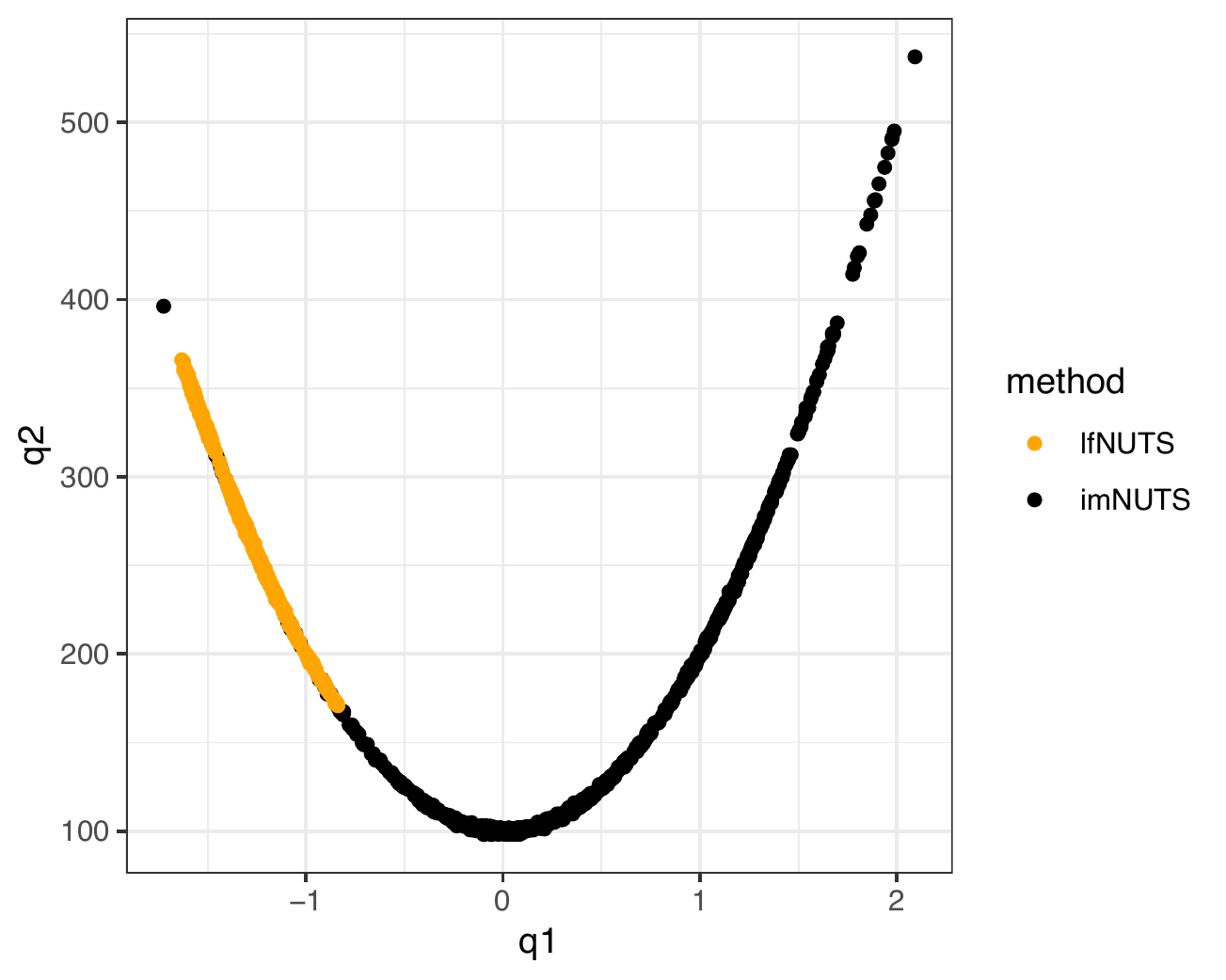}
\caption{1,000 samples of the banana distribution from lfNUTS and iNUTS. Because of the multiscale nature of the Hamiltonian system that arises when sampling the banana distribution, lfNUTS is forced to use an unreasonably small stepsize that  nearly reduces its behavior to that of a random walk and renders it unable to explore the entire distribution (orange samples). Meanwhile, iNUTS can effectively take ten times as large of a stepsize which allows it to efficiently explore the entire posterior.}
\label{fig:banana}       % Give a unique label
\end{figure}

The highly-nonlinear correlation stucture of this distribution is typical for complicated non-Gaussian distributions. For HMC, the long direction across the length of the banana compared to its narrow width creates a prototypical multiscale problem, forcing leapfrog to take an excessively small stepsize that leads to insufficient exploration of the posterior in lfNUTS compared to iNUTS (Figure \ref{fig:banana}). 

%

%%%%%%%%%%%%%%%%%%%%%%%%%
%%%%%%%%%%%%%%%%%%%%%%%%%
\subsection{Neal's Funnel}
Neal's $N+1$-dimensional funnel is a highly non-Gaussian distribution over $q \in \mathbb{R}^{N+1}$ defined as follows \citep{neal2003slice}:

\begin{eqnarray}
q_1 &\sim& \mathcal{N}(0, 3),\, i = 1\\ 
q_i &\sim& \mathcal{N}(0, e^{-q_1}),\, 2 \le i \le N+1.
\end{eqnarray}

\begin{figure}
  \includegraphics[width=0.47\textwidth]{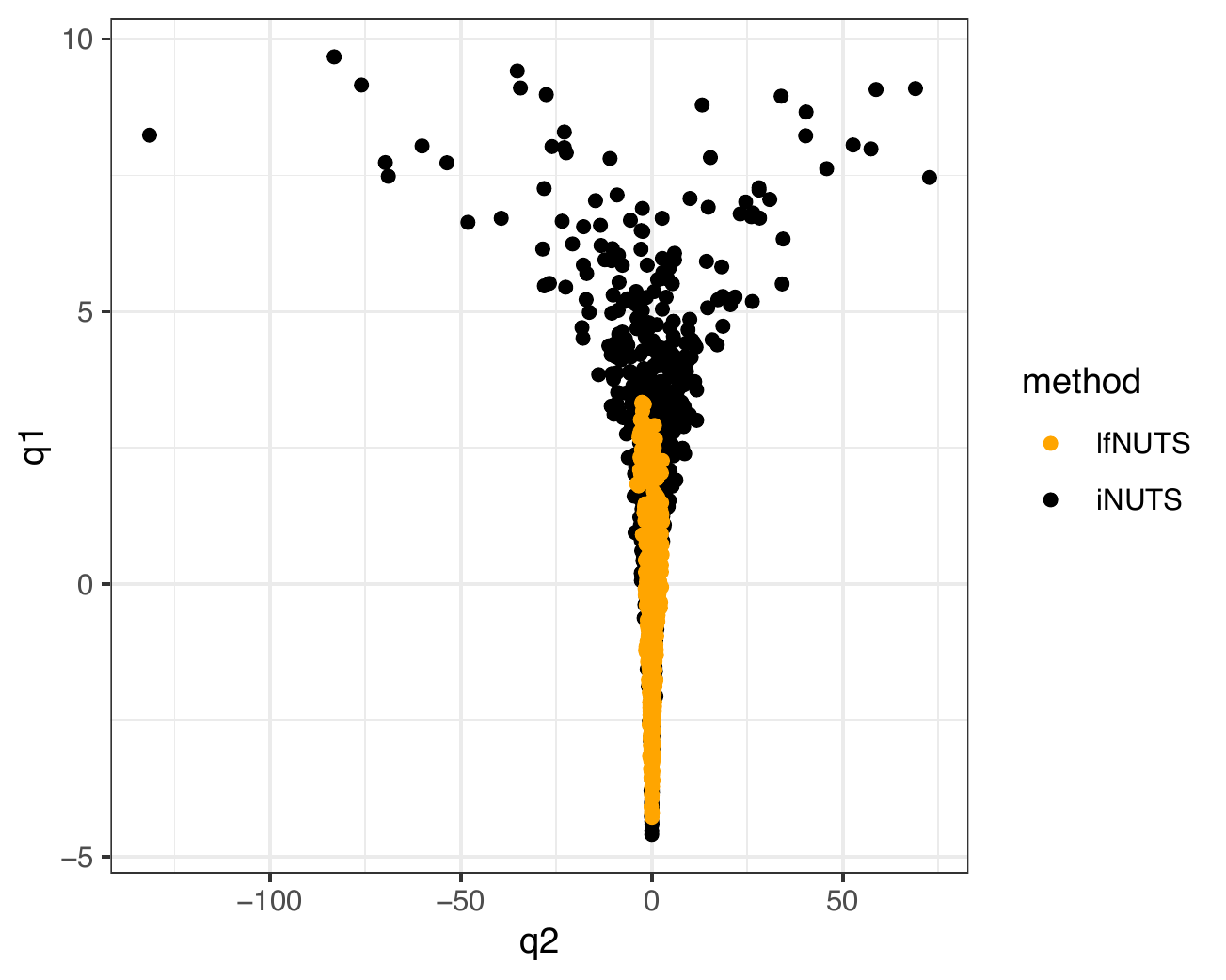}
\caption{1,000 samples of the 11-dimensional Neal's funnel distribution using lfNUTS and iNUTS. The curvature of the funnel varies drastically, at times being quite multiscale in nature (neck of the funnel) and at other times being more even in nature (mouth of the funnel). The multiscale regime places an unreasonably small stepsize on lfNUTS, causing the sampler to fail to efficiently explore the entire distribution in a reasonable amount of time. Meanwhile, iNUTS is able to maintain a reasonable stepsize and can sample efficiently.}
\label{fig:funnel}       % Give a unique label
\end{figure}

This is a particularly important distribution as it captures the characteristics of typical Bayesian hierarchical models that are known to give any sampler difficulty. In particular, the potential energy surface of the distribution exhibits a multiscale nature that changes drastically as a function of $q_1$, causing trouble for the lfNUTS compared to iNUTS. To quantify this difference, we took 1,000 samples of an 11-dimensional Neal's funnel using both lfNUTS and iNUTS. The scatter plot in Figure \ref{fig:funnel}, which depicts these samples for the first two dimensions, illustrates the widely-varying curvature of the distribution. For our experiments we used an identity mass matrix and picked the leapfrog stepsize according to the hand-tuned value of $h=0.003$ reported in \cite{betancourt2013general}. For the implicit midpoint stepsize we picked the largest stepsize that did not produce Newton convergence errors during sampling, which was $h = 0.2$.

A summary of results comparing the two methods is presented in Table 1. Because the implicit midpoint steps within iNUTS require a nonlinear solve, there is much more overhead in each step compared to leapfrog. However, because implicit midpoint has a much larger stability limit than leapfrog, iNUTS is able to take a much larger step, leading to more efficient samples. In practice, iNUTS ends up needing to do only a third of the work as lfNUTS to obtain an effective sample. In terms of computer time, iNUTS took about $30\%$ longer to obtain over ten times as many effective samples.

\begin{table}
\label{tab:funel}       % Give a unique label
\begin{tabular}{lll}
\hline\noalign{\smallskip}
 & lfNUTS & iNUTS  \\
\noalign{\smallskip}\hline\noalign{\smallskip}
avg. grad evals/step & 2.00 & 11.98\\
avg. hess-vec evals/step & 0.00 & 9.88 \\
avg. work/step & 2.00 & 21.86\\
avg. tree depth & 9.97 & 5.40 \\
avg. effective samples & 42.47 & 613.26 \\
work/effective sample & 47.09 & 16.11 \\
total computer time (s) & 181 & 231 \\
\noalign{\smallskip}\hline
\end{tabular}
\caption{A comparison of the computation involved in the sampling experiment. Average gradient evaluations and Hessian-vector evaluations are per step of the integrator, and work is the sum of these. Average effective samples are over all 11 dimensions being sampled. While iNUTS requires more work per step, because it is able to take a much larger step than leapfrog, it can obtain many more effective samples in a similar time frame.}
\end{table}

%%%%%%%%%%%%%%%%%%%%%%%%%
%%%%%%%%%%%%%%%%%%%%%%%%%
%\subsection{Poisson Gaussian Process}

%%%%%%%%%%%%%%%%%%%%%%%%%
%%%%%%%%%%%%%%%%%%%%%%%%%
%%%%%%%%%%%%%%%%%%%%%%%%%
\section{Discussion}
\label{discussion}
We have shown that distributions exhibiting multiple scales limit the stepsize of the leapfrog integrator in HMC, and how this limitation can be effectively circumvented by utilizing the implicit midpoint integrator together with Newton-Krylov iteration. Furthermore, we offered a practical implementation of implicit midpoint that is applicable to Bayesian posterior sampling problems, and provided practical guidelines for choosing which integrator to use, as well as the stepsize to use in the integrator. As illustrated in our examples, using implicit midpoint together with Newton-Krylov instead of leapfrog can provide a practical and significant efficiency boost in the context of NUTS.

%While our work provides a practical solution to the inefficiency of leapfrog-based HMC on multiscale distributions, we believe there is still much much to be gained by applying ideas from numerical integration to HMC and related methods. 

%%%%%%%%%%%%%%%%%%%%%%%%%
%%%%%%%%%%%%%%%%%%%%%%%%%
%%%%%%%%%%%%%%%%%%%%%%%%%
\appendix
\section{Eigenvalues of the Update Matrix for Implicit Midpoint on a Linear System}
\label{appendix_implicit_eigenvalues}
For the simple univariate Gaussian system with mass matrix one defined in Section \ref{numerical_stability}, the update equations (\ref{eq:midpoint_update_equations}) reduce to

\begin{eqnarray}
\label{eq:midpoint_update_equations_gaussian}
q_{n+1} &=& q_n +  h \left( \frac{p_n + p_{n+1}}{2} \right)   \\
p_{n+1} &=& p_n - h \sigma^2  \left( \frac{q_n + q_{n+1}}{2} \right) \nonumber.
\end{eqnarray}

\noindent In matrix notation, these update equations can be rewritten as

\begin{eqnarray}
\begin{pmatrix}
q_{n+1}\\
p_{n+1}
\end{pmatrix}
= 
\begin{pmatrix}
1 & \frac{h}{2}\\
-\frac{h}{2}\sigma^2 & 1
\end{pmatrix}
\begin{pmatrix}
q_n\\
p_n
\end{pmatrix}
+
\begin{pmatrix}
0 & \frac{h}{2}\\
-\frac{h}{2}\sigma^2 & 0
\end{pmatrix}
\begin{pmatrix}
q_{n+1}\\
p_{n+1}
\end{pmatrix}.
\end{eqnarray}

\noindent Collecting the $n+1$ terms on the right yields

\begin{eqnarray}
\begin{pmatrix}
1 & -\frac{h}{2}\\
\frac{h}{2}\sigma^2 & 1
\end{pmatrix}
\begin{pmatrix}
q_{n+1}\\
p_{n+1}
\end{pmatrix}
= 
\begin{pmatrix}
1 & \frac{h}{2}\\
-\frac{h}{2}\sigma^2 & 1
\end{pmatrix}
\begin{pmatrix}
q_n\\
p_n
\end{pmatrix}
\end{eqnarray}

\noindent which yields the update equation

\begin{eqnarray}
\label{eq:midpoint_linear_update_matrix}
\begin{pmatrix}
q_{n+1}\\
p_{n+1}
\end{pmatrix}
&=& 
\begin{pmatrix}
1 & -\frac{h}{2}\\
\frac{h}{2}\sigma^2 & 1
\end{pmatrix}^{-1}
\begin{pmatrix}
1 & \frac{h}{2}\\
-\frac{h}{2}\sigma^2 & 1
\end{pmatrix}
\begin{pmatrix}
q_n\\
p_n
\end{pmatrix}
\nonumber
\\
&=&
\frac{1}{1 + \frac{h^2}{4} \sigma^2}
\begin{pmatrix}
1 & \frac{h}{2}\\
-\frac{h}{2}\sigma^2 & 1
\end{pmatrix}
\begin{pmatrix}
1 & \frac{h}{2}\\
-\frac{h}{2}\sigma^2 & 1
\end{pmatrix}
\begin{pmatrix}
q_n\\
p_n
\end{pmatrix}
\nonumber
\\
&=&
\frac{1}{1 + \frac{h^2}{4} \sigma^2}
\begin{pmatrix}
1 - \frac{h^2}{4} \sigma^2 & h\\
-h \sigma^2 & 1-\frac{h^2}{4} \sigma^2
\end{pmatrix}
\begin{pmatrix}
q_n\\
p_n
\end{pmatrix}.
\end{eqnarray}

\noindent Note that the eigenvalues of the scalar times the matrix in this equation are simply the eigenvalues of the matrix times the scalar in front. Defining $\mu := (h^2/4) \sigma^2$ these eigenvalues are the solutions of the quadratic equation

\begin{equation}
\lambda^2 - 2(1-\mu) + (1-\mu)^2 + h^2 \sigma^2 =0
\end{equation}

\noindent which by the quadratic formula are 

\begin{equation}
\lambda_{1,2} = (1-\mu) \pm h\sigma i.
\end{equation}

\noindent Thus the moduli of the eigenvalues of the overall update matrix are one.

\begin{acknowledgements}
Funding for this project was provided by the U.S. Army Research Office under Coagulopathy grant W911NF-10-2-0114.
\end{acknowledgements}

\bibliographystyle{plainnat}
\bibliography{implicit}   % name your BibTeX data base

\begin{thebibliography}{27}
\providecommand{\natexlab}[1]{#1}
\providecommand{\url}[1]{\texttt{#1}}
\expandafter\ifx\csname urlstyle\endcsname\relax
  \providecommand{\doi}[1]{doi: #1}\else
  \providecommand{\doi}{doi: \begingroup \urlstyle{rm}\Url}\fi

\bibitem[Ascher and Petzold(1998)]{ascher1998computer}
Uri~M Ascher and Linda~R Petzold.
\newblock \emph{Computer methods for ordinary differential equations and
  differential-algebraic equations}, volume~61.
\newblock SIAM, 1998.

\bibitem[Ascher and Reich(1999)]{ascher1999some}
Uri~M Ascher and Sebastian Reich.
\newblock On some difficulties in integrating highly oscillatory {Hamiltonian}
  systems.
\newblock In \emph{Computational Molecular Dynamics: Challenges, Methods,
  Ideas}, pages 281--296. Springer, 1999.

\bibitem[Beskos et~al.(2013)Beskos, Pillai, Roberts, Sanz-Serna, Stuart,
  et~al.]{beskos2013optimal}
Alexandros Beskos, Natesh Pillai, Gareth Roberts, Jesus-Maria Sanz-Serna,
  Andrew Stuart, et~al.
\newblock Optimal tuning of the hybrid {Monte} {Carlo} algorithm.
\newblock \emph{Bernoulli}, 19\penalty0 (5A):\penalty0 1501--1534, 2013.

\bibitem[Betancourt(2013)]{betancourt2013general}
Michael Betancourt.
\newblock A general metric for {Riemannian} manifold {Hamiltonian} {Monte}
  {Carlo}.
\newblock In \emph{Geometric science of information}, pages 327--334. Springer,
  2013.

\bibitem[Betancourt(2017)]{betancourt2017conceptual}
Michael Betancourt.
\newblock A conceptual introduction to {Hamiltonian} {Monte} {Carlo}.
\newblock \emph{arXiv preprint arXiv:1701.02434}, 2017.

\bibitem[Carpenter et~al.(2017)Carpenter, Gelman, Hoffman, Lee, Goodrich,
  Betancourt, Brubaker, Guo, Li, and Riddell]{carpenter2017stan}
Bob Carpenter, Andrew Gelman, Matthew~D Hoffman, Daniel Lee, Ben Goodrich,
  Michael Betancourt, Marcus Brubaker, Jiqiang Guo, Peter Li, and Allen
  Riddell.
\newblock Stan: A probabilistic programming language.
\newblock \emph{Journal of {Statistical} {Software}}, 76\penalty0 (1), 2017.

\bibitem[Chao et~al.(2015)Chao, Solomon, Michels, and Sha]{chao2015exponential}
Wei-Lun Chao, Justin Solomon, Dominik Michels, and Fei Sha.
\newblock Exponential integration for {Hamiltonian} {Monte} {Carlo}.
\newblock In \emph{International Conference on Machine Learning}, pages
  1142--1151, 2015.

\bibitem[Christensen et~al.(2005)Christensen, Roberts, and
  Rosenthal]{christensen2005scaling}
Ole~F Christensen, Gareth~O Roberts, and Jeffrey~S Rosenthal.
\newblock Scaling limits for the transient phase of local metropolis--hastings
  algorithms.
\newblock \emph{Journal of the Royal Statistical Society: Series B (Statistical
  Methodology)}, 67\penalty0 (2):\penalty0 253--268, 2005.

\bibitem[Duane et~al.(1987)Duane, Kennedy, Pendleton, and
  Roweth]{duane1987hybrid}
Simon Duane, Anthony~D Kennedy, Brian~J Pendleton, and Duncan Roweth.
\newblock Hybrid {Monte} {Carlo}.
\newblock \emph{Physics {Letters} B}, 195\penalty0 (2):\penalty0 216--222,
  1987.

\bibitem[Eisenstat and Walker(1996)]{eisenstat1996choosing}
Stanley~C Eisenstat and Homer~F Walker.
\newblock Choosing the forcing terms in an inexact newton method.
\newblock \emph{SIAM Journal on Scientific Computing}, 17\penalty0
  (1):\penalty0 16--32, 1996.

\bibitem[Gelman et~al.(2013)Gelman, Stern, Carlin, Dunson, Vehtari, and
  Rubin]{gelman2013bayesian}
Andrew Gelman, Hal~S Stern, John~B Carlin, David~B Dunson, Aki Vehtari, and
  Donald~B Rubin.
\newblock \emph{Bayesian {Data} {Analysis}}.
\newblock Chapman and Hall/CRC, 2013.

\bibitem[Girolami and Calderhead(2011)]{girolami2011riemann}
Mark Girolami and Ben Calderhead.
\newblock Riemann manifold langevin and hamiltonian monte carlo methods.
\newblock \emph{Journal of the Royal Statistical Society: Series B (Statistical
  Methodology)}, 73\penalty0 (2):\penalty0 123--214, 2011.

\bibitem[Hairer et~al.(2006)Hairer, Lubich, and Wanner]{hairer2006geometric}
Ernst Hairer, Christian Lubich, and Gerhard Wanner.
\newblock \emph{Geometric numerical integration: structure-preserving
  algorithms for ordinary differential equations}, volume~31.
\newblock Springer Science \& Business Media, 2006.

\bibitem[Hoffman and Gelman(2014)]{hoffman2014no}
Matthew~D Hoffman and Andrew Gelman.
\newblock The no-u-turn sampler: adaptively setting path lengths in hamiltonian
  monte carlo.
\newblock \emph{Journal of Machine Learning Research}, 15\penalty0
  (1):\penalty0 1593--1623, 2014.

\bibitem[Knoll and Keyes(2004)]{knoll2004jacobian}
Dana~A Knoll and David~E Keyes.
\newblock Jacobian-free {Newton}--{Krylov} methods: a survey of approaches and
  applications.
\newblock \emph{Journal of Computational Physics}, 193\penalty0 (2):\penalty0
  357--397, 2004.

\bibitem[Leimkuhler and Matthews(2016)]{leimkuhler2016molecular}
Ben Leimkuhler and Charles Matthews.
\newblock \emph{Molecular {Dyamics}.}
\newblock Springer, 2016.

\bibitem[Leimkuhler and Reich(2004)]{leimkuhler2004simulating}
Benedict Leimkuhler and Sebastian Reich.
\newblock \emph{Simulating hamiltonian dynamics}, volume~14.
\newblock Cambridge {University} press, 2004.

\bibitem[Long et~al.(2013)Long, Wolfe, Mashner, and Chirikjian]{long2013banana}
Andrew~W Long, Kevin~C Wolfe, Michael~J Mashner, and Gregory~S Chirikjian.
\newblock The banana distribution is {Gaussian}: A localization study with
  exponential coordinates.
\newblock \emph{Robotics: Science and Systems VIII; MIT Press: Cambridge, MA,
  USA}, pages 265--272, 2013.

\bibitem[Metropolis et~al.(1953)Metropolis, Rosenbluth, Rosenbluth, Teller, and
  Teller]{metropolis1953equation}
Nicholas Metropolis, Arianna~W Rosenbluth, Marshall~N Rosenbluth, Augusta~H
  Teller, and Edward Teller.
\newblock Equation of state calculations by fast computing machines.
\newblock \emph{The {Journal} of {Chemical} {Physics}}, 21\penalty0
  (6):\penalty0 1087--1092, 1953.

\bibitem[Meyer(2000)]{meyer2000matrix}
Carl~D Meyer.
\newblock \emph{Matrix {Analysis} and {Applied} {Linear} {Algebra}}, volume~71.
\newblock SIAM, 2000.

\bibitem[Neal(2003)]{neal2003slice}
Radford~M Neal.
\newblock Slice sampling.
\newblock \emph{Annals of statistics}, pages 705--741, 2003.

\bibitem[Neal et~al.(2011)]{neal2011mcmc}
Radford~M Neal et~al.
\newblock {MCMC} using {Hamiltonian} dynamics.
\newblock \emph{Handbook of Markov Chain Monte Carlo}, 2\penalty0
  (11):\penalty0 2, 2011.

\bibitem[Okudo and Suzuki(2015)]{okudo2015hamiltonian}
Michiko Okudo and Hideyuki Suzuki.
\newblock Hamiltonian {Monte} {Carlo} with explicit, reversible, and
  volume-preserving adaptive step size control.
\newblock \emph{JSIAM Letters}, 9:\penalty0 33--36, 2015.

\bibitem[Quarteroni et~al.(2010)Quarteroni, Sacco, and
  Saleri]{quarteroni2010numerical}
Alfio Quarteroni, Riccardo Sacco, and Fausto Saleri.
\newblock \emph{Numerical {Mathematics}}, volume~37.
\newblock Springer Science \& Business Media, 2010.

\bibitem[Roberts and Stramer(2002)]{roberts2002langevin}
Gareth~O Roberts and Osnat Stramer.
\newblock Langevin diffusions and metropolis-hastings algorithms.
\newblock \emph{Methodology and computing in applied probability}, 4\penalty0
  (4):\penalty0 337--357, 2002.

\bibitem[Saad and Schultz(1986)]{saad1986gmres}
Youcef Saad and Martin~H Schultz.
\newblock {GMRES}: A generalized minimal residual algorithm for solving
  nonsymmetric linear systems.
\newblock \emph{SIAM Journal on {Scientific} and {Statistical} {Computing}},
  7\penalty0 (3):\penalty0 856--869, 1986.

\bibitem[Shahbaba et~al.(2014)Shahbaba, Lan, Johnson, and
  Neal]{shahbaba2014split}
Babak Shahbaba, Shiwei Lan, Wesley~O Johnson, and Radford~M Neal.
\newblock Split {Hamiltonian} {Monte} {Carlo}.
\newblock \emph{Statistics and Computing}, 24\penalty0 (3):\penalty0 339--349,
  2014.

\end{thebibliography}

%\begin{thebibliography}{}
% etc
%\end{thebibliography}

\end{document}